# Electrochemical polymerization of anilinium hydrochloride

Yomen Atassi[1] and Mohammad Tally

Department of Applied physics, Higher Institute for Applied Sciences and Technology, P.O. Box 31983, Damascus, Syria.

**Abstract:** Electropolymerization of anilinium hydrochloride was carried out on platinum plate electrode in a protic medium by cyclic voltammetry. The effects on the electrodeposition of the sweep rate, monomer concentration, pH of the medium and electrode nature are discussed.

[1] E-mail address: yomen.atassi@hiast.edu.sy

# 1. Introduction

A polymeric material that possesses the electronic, magnetic, electrical and optical properties of a metal while retaining the processability and mechanical properties usually associated with a conventional organic polymer is called a *conducting polymer.*

It has been recognized that any organic polymer containing a conjugated $\pi$ – electron system can always have metallic conductivity upon addition of electrons or holes by doping (chemically or electrochemically). Metallic conductivities have been observed in a number of conjugated systems, for example, polyaniline, polypyrrole, polythiophene and poly(p-phenylen). In addition to their use as electrical conductors, conjugated polymers are becoming more and more important as functional materials in other areas, such as optics (transparent/flexible electrodes and electrochromic devices), nonlinear optics, microelectronics (light-emitting diodes and organic transistors), microelectromechanical systems (sensors, rechargeable batteries, and artificial muscles), electromagnetic devices (static charge dissipating and electromagnetic shielding), and membranes for selective gas separation. Many of these concepts are in various stages of development, and some commercial products such as rechargeable lithium-polymer batteries have been marketed by Seiko, BASF, and Allied Signal.

Polyaniline (PANI) is probably the oldest organic polymer ever synthesized (in 1862) [1]. In the early work published in 1910s, it was described as existing in four different oxidation states; each of them was an octomer. In 1967, Josefowicz et al. [2] reported that the conductivity of polyaniline increases by several orders of magnitude when doped in protonic acids with decreasing pH values. Tjey also recognized that polyaniline could serve as an electrode material for rechargeable batteries. Unfortunately, all these early studies were fraught with problems such as uncertain composition and were lost in literature. In 1980 Diaz and Logan [3] reactivated research on polyanilines and several other groups immediately followed up during the following decades [4].

Although (PANI) can be prepared by chemical means, the most widely used technique is electrochemical anodic oxidation. It presents several advantages, such as absence of catalyst, direct grafting of the doped conducting polymer onto the electrode surface, and easy control of the film thickness by integration of deposition charge. Monomer oxidation must occur in a solvent system characterized by a suitable potential window, and the species produced must react preferentially to form the polymer. Relatively high monomer concentrations (typically 0.1M) in an acidic medium are generally used, as well as inert electrodes.

Cyclic voltammetry (CV) is an excellent technique for the investigation of the best operative conditions. During the first potential scan, the current maximum corresponds to the monomer oxidation potential. Continuous cycling of the potential over the peak develops the redox cycle of the growing electroactive PANI. Reversible oxidation of the polymers (doping-dedoping process) occurs at potentials lower than that of the corresponding monomer and, therefore, the process is put in evidence during potential scan.

The mechanism generally accepted for the electropolymerization of aniline by anodic coupling is shown in scheme 1.
This mechanism is repeated for every cycle of sweeping and it allows interpreting the augmentation of the concentration of the oxidized specie of the quasi-reversible couple ($i_p$ increases).
The first step is the formation of an aniline radical cation. This radical aniline is resonance stabilized as shown in step two of the mechanism. These radicals may then combine in different ways to form polyaniline. Head-to-tail coupling or 1-, 4- substitution is believed to predominate and produce p-aminodiphenyl amine in step three of the mechanism.

We have to notice that the proposed mechanism is the precursor of another one more complex which explains the polyaniline film formation.

In the present paper, we report electropolymerization from an electrolyte containing anilinium hydrochloride monomer in an acidified medium. The polymer films have been studied by cyclic voltammetry. In fact this method reveals qualitatively the reversibility of electron transfer during the electropolymerization and also examines the electroactivity of the polymer film because the oxidation and reduction can be monitored in the form of a current-potential diagram.

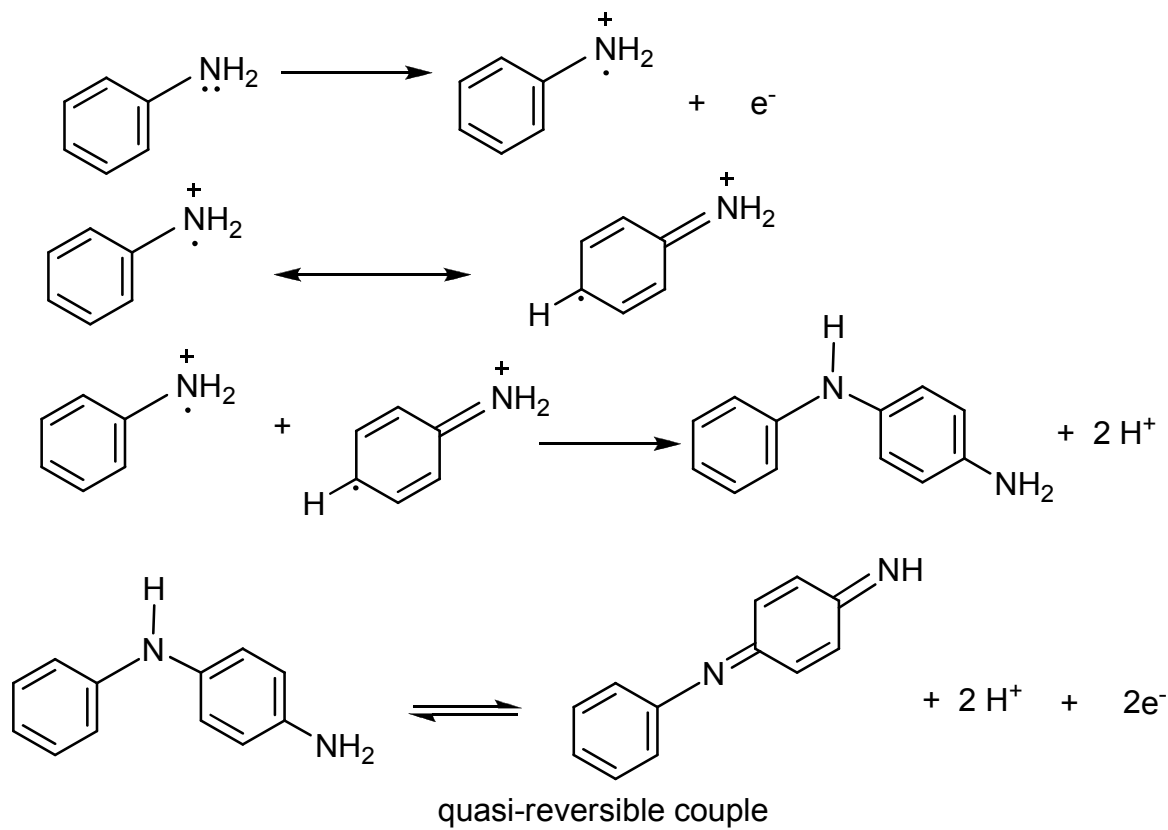

Scheme 1. Mechanism of oxidative electropolymerization of aniline.

## 2. Materials and methods:

A Voltalab potentiostat-galvanostat was used in the experiments. Electrochemical experiments were performed in a three electrode cell, containing a platinum plate (1cm×1cm) working electrode, a platinum wire auxiliary electrode and a KCl saturated calomel electrode (SCE). The potential of the SCE with respect to the standard electrode of hydrogen (SHE) is 0.247 mV at $T = 25°C$.

We used two types of electrolytes:
- A solution of anilinium hydrochloride 0.4M in water, acidified with concentrated HCl.
- A solution of aniline 0.1M in aqueous sulfuric acid 1M.

Aniline (Merck) was distilled under reduced pressure and stored in darkness before use. As received sulfuric acid (Prolabo), anilinium hydrochloride (Merck) and distilled water were used in preparing the electrolyte solution.

## 3. Results and discussions:

3.1. **Film preparation:**

3.1.1. *Electrodeposition*:

The solution was deoxygenated by purging with $N_2$ for about 15 min and maintaining $N_2$ current over the solution all over the experiment. In fact the soluble oxygen in the solution constitutes electroactive specie, and its reduction may interfere with the electropolymerization.

Fig. (1) shows a typical cyclic voltammetry CV of aniline polymerized on platinum at a scan rate of 100mV/s.

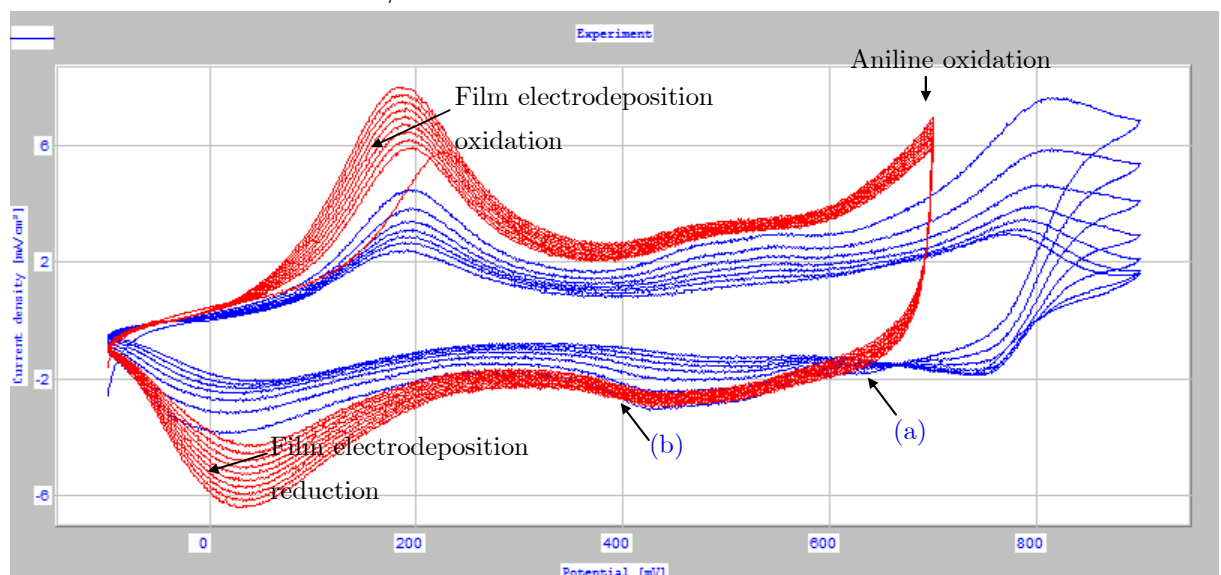

Figure 1: (a): 7 Cycles from -100mV to 900 mV and (b): 10 Cycles from -100mV to 700 mV at a scan rate of 100 mV/s of 0.1M aniline solution in 1M $H_2SO_4$.

At first, we record the variation of the current density while the potential varies between -100 mV and +900 mV (versus SCE), Fig. 1 (a). On the first scan an oxidation peak of aniline can be seen at about 800 mV. It's necessary to reach this potential in order to start the synthesis of the polymeric film. We also notice an oxidation peak at about 190 mV vs. SCE, with a reverse peak at about 80 mV vs. SCE.

On subsequent sweeps this oxidation peak shifts in an anodic direction and the reduction peak shifts in a cathodic direction with an increase in current for both peaks. The electrochemical behavior indicates that the conducting polymer is being formed, and a polymer deposit can be clearly seen.

After 7 cycles of scanning, we diminish the value of the potential to 700 mV and we record 10 cycles, otherwise we favor the degradation of the polymeric film, Fig. 1, (b).

The deposition yielded a deep green film of polyaniline, Fig. 2.

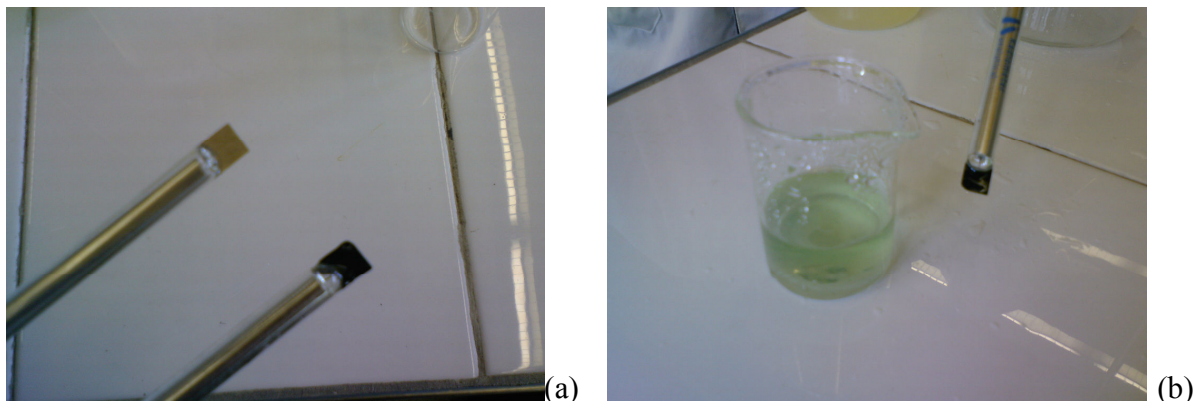

Figure 2: (a) 2-electrodes of platinum, one with a deposited PANI film and the other without. (b) Dissolution of the film in chloroform and the formation of a green solution of PANI.

The deposited film exhibits a smooth surface, uniform distribution and good adhesion to the electrode surface.

At first, aniline oxidation is not a reversible phenomena, as the signal corresponding to this has no corresponding one while we perform the reverse sweeping, This implies that the reduction of the resultant product is not possible. Since the product of aniline oxidation is engaged in another chemical reaction (proposed mechanism), we can assume that the reduction does not occur.

After few minutes, we notice the apparition of a signal in a region that does not present a peak before. This proves that new electroactive specie is now present in the medium. It corresponds to a quasi-reversible system (scheme 1).

The film deposited on platinum electrode is polyaniline.

Sometimes, if the aniline is sufficiently pure and the platinum electrode is clean, we can observe the film deposition at the beginning of the experiment.

As in usual solution-based CV, a triangular-shaped potential is applied to the cell, but in this case the working electrode is coated with the polymer to be studied. When current is flowing there is electron transfer across the metal-polymer interface and simultaneously ion transfer across the polymer-solution interface. The only diffusion-controlled process occurs inside the polymer film, where ions have limited mobility. If the polymer film is very thin, the diffusion time of ions is very short and we expect that the reverse electron transfer occurs exactly at the same potential on the return sweep of CV; i.e., we should have a voltammogram with symmetrical and mirror-image cathodic and anodic waves.

From the cyclic voltammogram it is possible to estimate the parameter $E^{0\prime}$, the formal standard potential by averaging the two peak potentials. This parameter is a useful indicator of the thermodynamic ease of oxidation. For polyaniline the value is -0.2V versus the standard calomel electrode (SCE).

The peak potential shifts regularly to more positive or more negative values with increasing thickness of the polymer film due to ohmic contribution to the overpotential.

3.1.2. *Electrochromism*:

The polymer can be switched between its reduced form, which is transparent and its oxidized form, which is dark green color, Fig. 3.

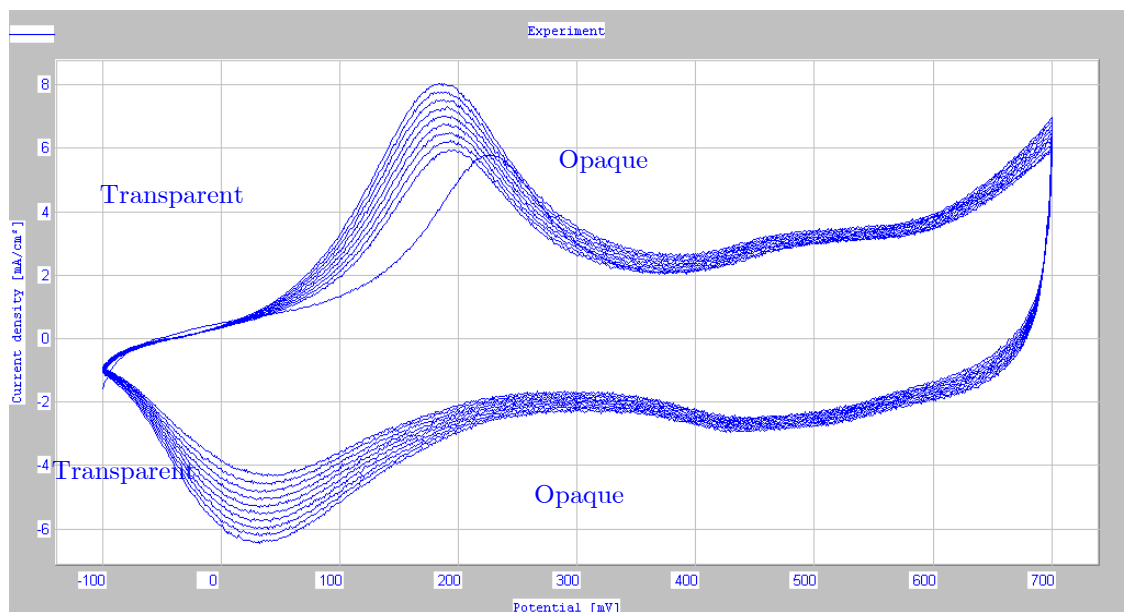

Figure 3: Switching between the transparent reduced form and the dark green oxidized form.

When the film is reduced (potential less than +100 mV), the polyaniline is electrically insulator and the film is optically transparent in the visible.

On the other hand, when the film is oxidized (potential more than +250 mV), the polyaniline is electrically conductor and the film is opaque (Fig.1).

In fact, the passage of polyaniline from reduced state which is insulator (leucoemeraldine) to the oxidized state which is conductor (protonated emeraldine) is accompanied with colour change (from incolor to green), scheme 2.

We can notice easily the color changes, by observing the platinum electrode which changes its color between transparent and dark green.

This electrochromic property of electrically conducting polymer have attracted great interest because of their potential application in energy-saving smart window or other light-modulation devices. Polyaniline is an interesting electrochromic polymer.

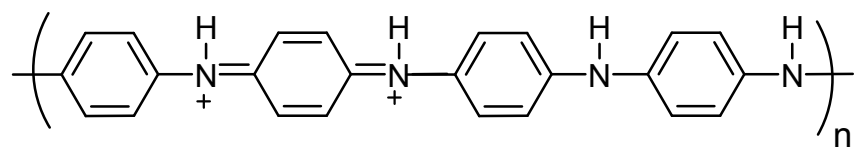

Emeraldine Salt , green, partially oxidized, protonated conducted.

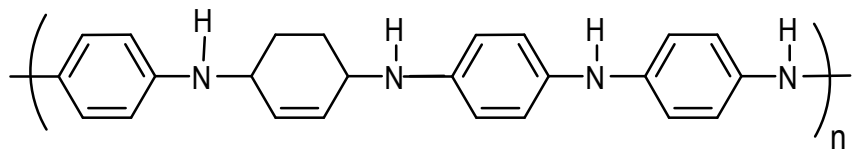

Leucoemeraldine, Clear, fully reduced, insulating.

Scheme 2. Emeraldine Salt and Leucoemeraldine

3.1.3. *Nucleation and film growth*:

The growth of the polymer film on the electrode surface starts from a number of separate sites, called nucleation sites or centers. The formation of nucleation sites is progressive, i.e. new centers are formed continuously during the growth process.

As for film growth, it is assumed that the growth of the polymer chains occurs in solution as a homogeneous reaction between oligomers and monomer cation radicals with subsequent precipitation of sparingly soluble oligomers onto the nucleation sites.

3.2. **Effect of monomer concentration:**

We record cyclic voltammograms of 2 electrolytes with 0.1M and 0.4M of monomer concentration in an acidified medium, Fig. 4.

In fact, as the increase of the redox wave currents implies that the amount of the polymer electrodeposited increased on the working electrode, we can consider the deposition current as a measure of polymerization rate.

By simple comparison between the intensities values, we find that the polymerization rate increases with monomer concentration (For a monomer concentration of 0.4M, we have $I = 28\,mA/cm^2$ of the $10^{th}$ cycle Fig. 4, (b) while for a monomer concentration of 0.1M, we have $I = 8\,mA/cm^2$ of the $10^{th}$ cycle, Fig.4, (a).

More precisely, Fig. 5 shows the change in anodic and cathodic density currents of polyaniline during the polymer growth for different monomer concentrations 0.4M and 0.1M at 100 mV/s. As the other experimental parameters were kept constant, the different slopes, dI/dt, obtained for different concentrations corresponds to the growth rate of polymer film on the platinum electrode. The ratio of slopes were 3.1:1 for anodic currents in 0.4M solution vs. 0.1M solution, and 3.6:1 for cathodic currents in 0.4M solution vs. 0.1M solution, indicating that electrochemical polymer growth rate ratio in anilinium is faster than in aniline.

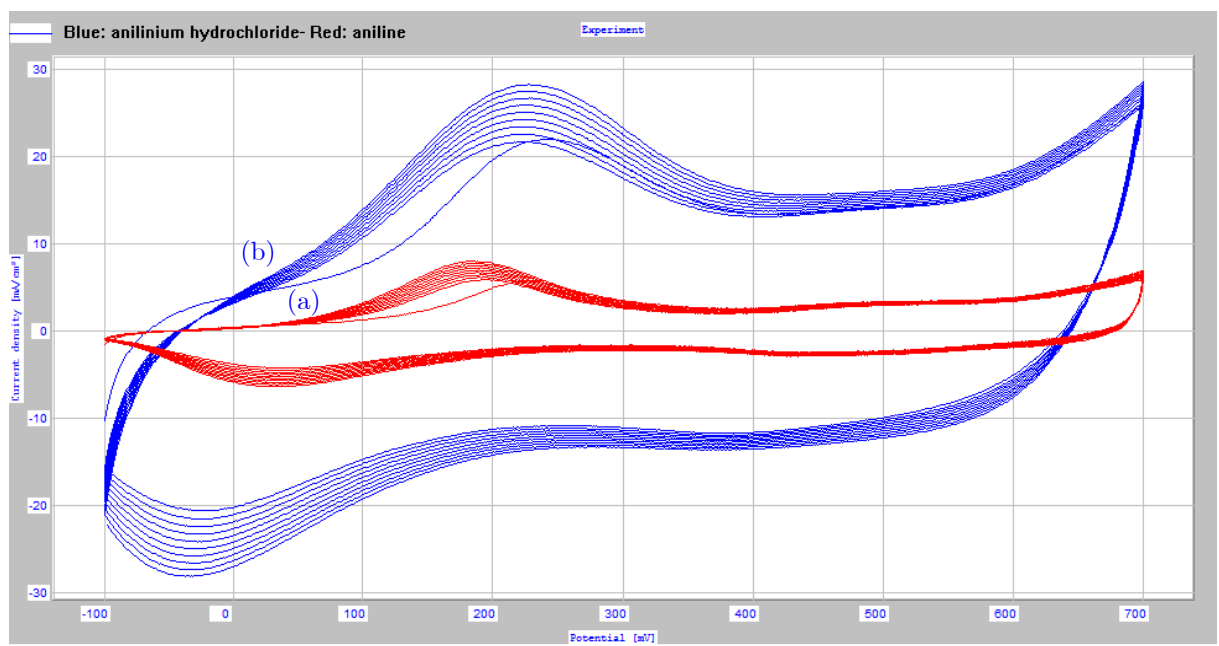

Figure 4: Effect of monomer concentration: (a) 0.1M and (b) 0.4M.

On the other hand, we note that high monomer concentration results in higher polymerization rate, with rough surface, flaky and poor adhesion film.

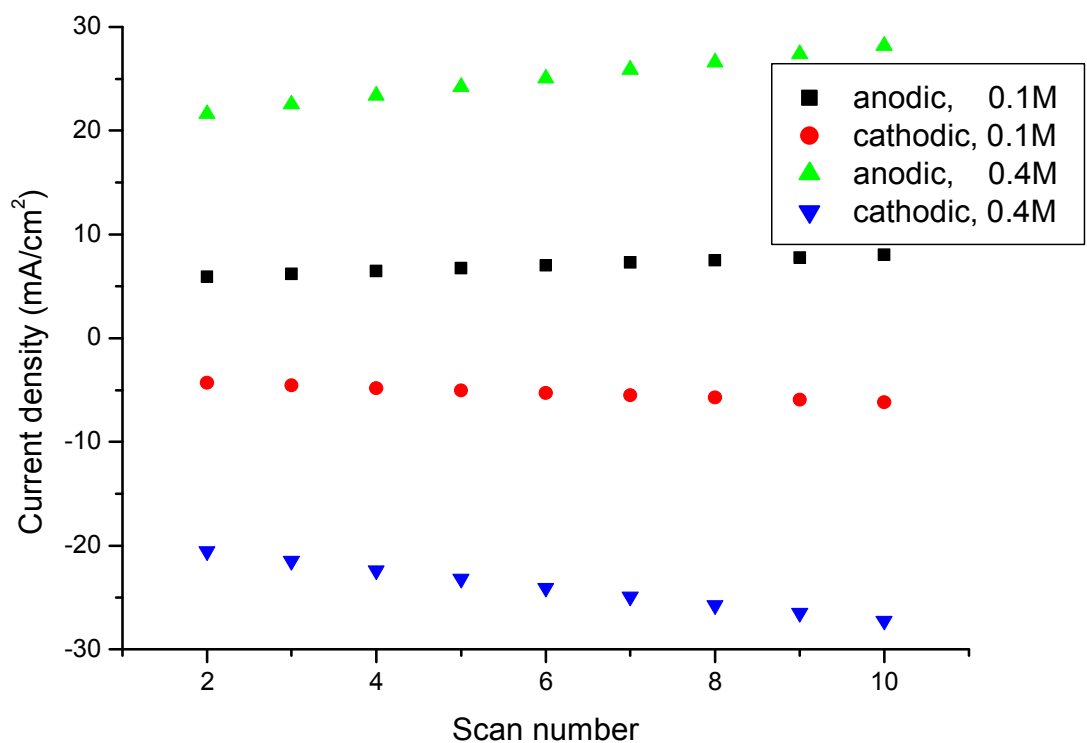

Figure 5: Anodic and cathodic peak density currents of polyaniline during polymer growth at 100 mV/s (anodic and cathodic peak currents were taken from CV at approximately 225 mV and -25 mV respectively for the 0.4 M monomer solution; and at approximately 190 mV and 35 mV respectively for the 0.1 M monomer solution).

**3.3 Effect of sweep rate:**

Fig. (6) represents CV of polyaniline formed by 5 cycles at (a) 50 mV/s, (b) 100 mV/s, (c) 150 mV/s, (d) 200 mV/s, (e) 250 mV/s, in 0.1 and 0.4 M aniline solutions.

The current density also increases/decreases in oxidation/reduction processes, more rapidly with the increase in sweep rate.

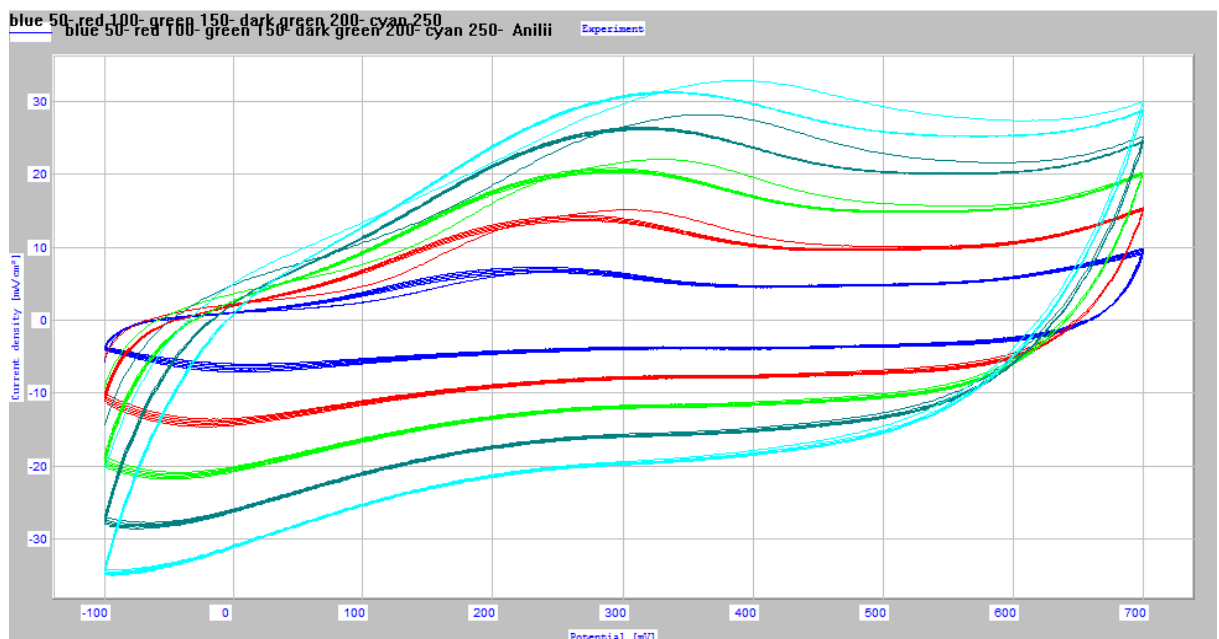

I-Anilinium hydrochloride solution

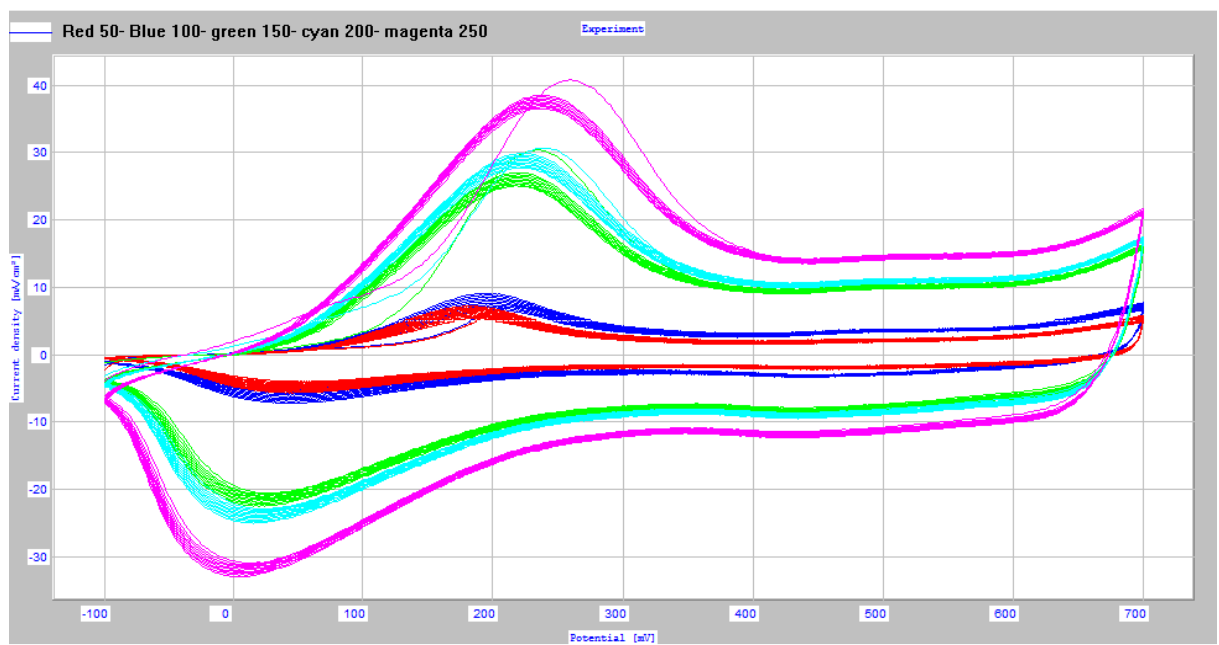

II-Aniline solution

Figure 6: The cyclic voltammograms of PANI versus SCE at different scan rates (from inner to outer: 50, 100, 150, 200, 250 mV/s) in I: 0.4M and II: 0.1 M monomer solutions.

As is typical for a redox reaction involving surface-attached species, the peak anodic current, $i_{pa}$ and the peak cathodic current, $i_{pc}$, scaled linearly with the sweep rate, Fig.(7).

An important point to signal here is that the current is directly proportional to the scan rate, whereas in conventional CV, where the transport of ions is limited by diffusion, the current is proportional to the square root of the scan rate.

The current in the reversible case is:

$$I = \frac{n^2 F^2 A \Gamma_T v}{4RT \cosh^2(\theta/2)}$$

Where $\theta = \frac{nF}{RT}(E - E°)$ and $\Gamma_T = \Gamma_{ox} + \Gamma_{red}$ is the total surface concentration of redox active specie, v is the scan rate, and other symbols have their usual electrochemical meaning.

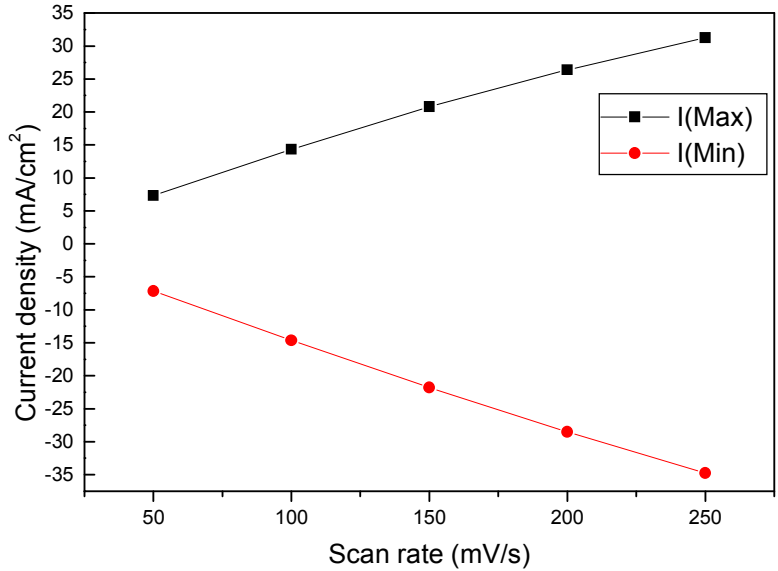

Figure 7: The peak anodic current, $i_{pa}$ and the peak cathodic current, $i_{pc}$, scaled linearly with the sweep rate.

### 3.4. Degradation of the film:

Figure 8 represents 7 Cycles from -100mV to 900 mV at a scan rate of 100 mV/s of 0.1M aniline solution in 1M $H_2SO_4$. As mentioned above, the oxidation peak of aniline can be seen at about 800 mV. It's necessary to reach this potential in order to start the synthesis of the polymeric film. But the peak of aniline oxidation rises with the number of cycles.

So, after a number of cycles of scanning, we have to diminish the value of the potential to 700 mV, otherwise we favor the degradation of the polymeric film.

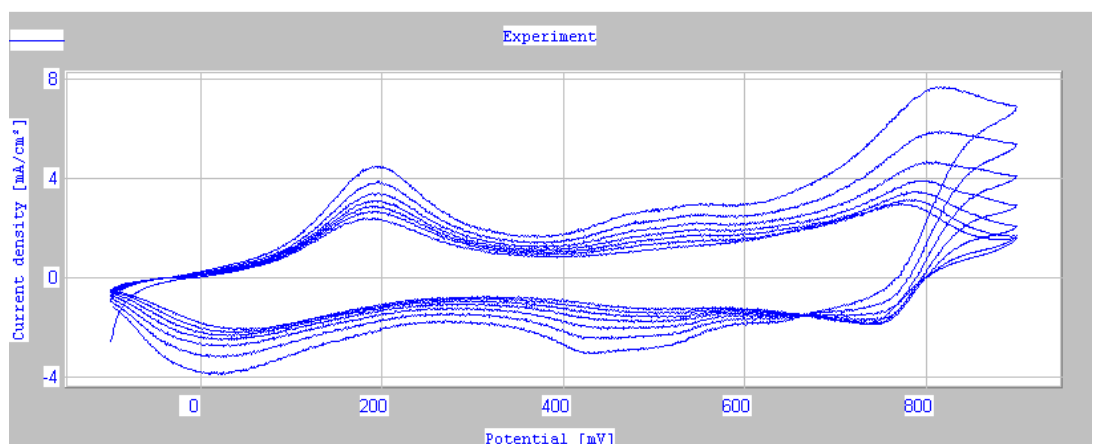

Figure 8: Seven cycles from -100mV to 900 mV at a scan rate of 100 mV/s of 0.1M aniline solution in 1M $H_2SO_4$.

### 3.5. Effect of working electrode:

We carried out an experiment using silver plate as a working electrode. We used the same electrodes as previously for the auxiliary and the reference ones: Pt wire and Calomel electrode respectively.

As it's clearly seen in Fig. (9), there's no formation of PANI film on silver electrode in the range -100 mV to 900mV.

So according to the nature of the working electrode, we have to choose the suitable potential window in order to deposit the polymeric film.

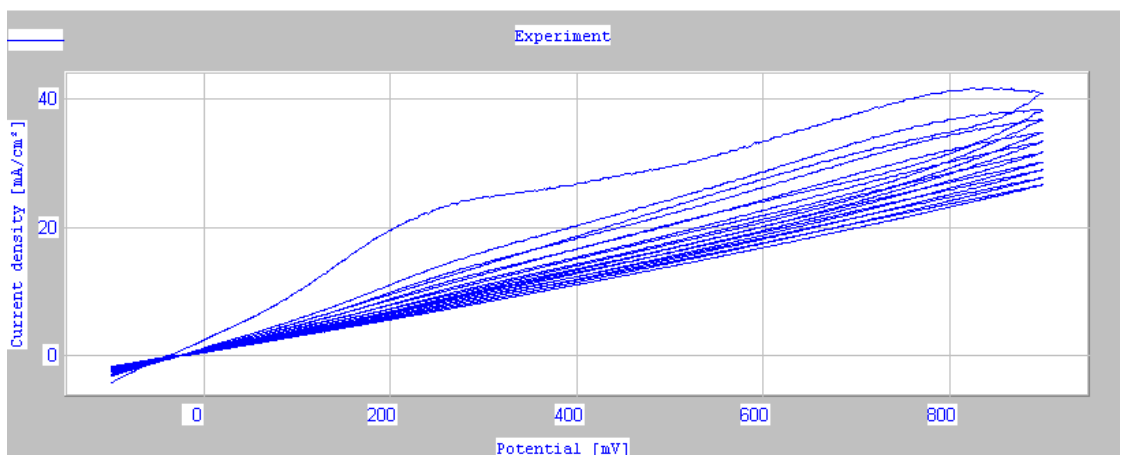
Figure 9: CV in the range -100 mV to 900mV using silver plate as working electrode.

### 3.6. Effect of acidic medium:

We carried out an experiment where no acid was added to the anilinium hydrochloride solution, Fig. 10.

We found that no film deposition had taken place in absence of an acidified medium. This indicates that a protic medium is a key factor in film preparation.
As it favors rapid protonation-deprotonation exchange giving rise to larger currents.

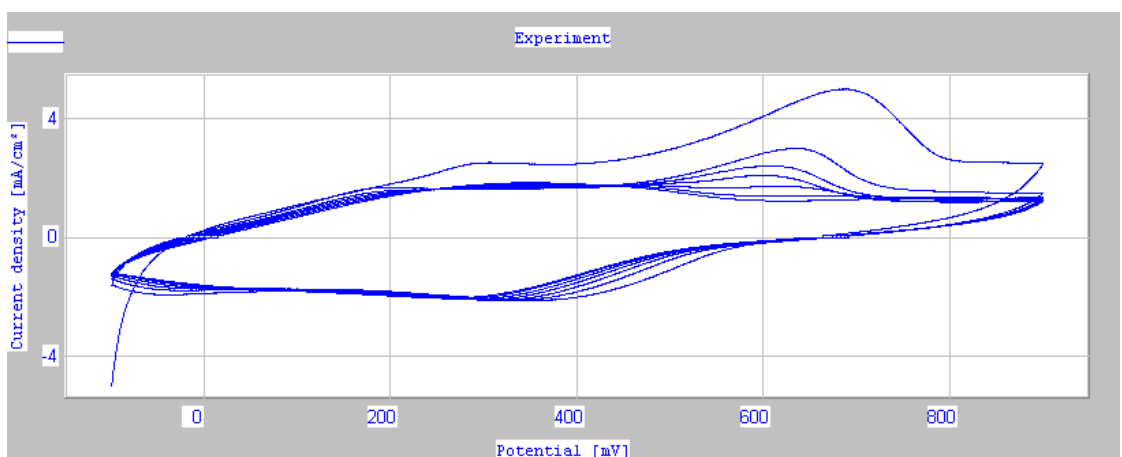
Figure 10: Part I : 7 Cycles, -100mV to 900 mV with a scan rate 100 mV/s
Solution of anilinium hydrochloride non acidified with concentrated HCl.

### 4. Conclusion:

In the present work polyaniline films have been deposited electrochemically.
Due to many applications of polyaniline, better control of its electropolymerization seems to be very important for the development of usable films. Simple modifications of the experimental parameters, e.g. changing the electrode material, monomer concentrations or sweep rate, result in changes in the electropolymerization process and in properties of the final film. The results show that for quality polyaniline films of smooth surface, better adhesion, less porosity and uniform distribution, the monomer concentration should be kept low (0.1 M), though the rate of polymerization is low at this concentration. Cyclic voltammetry results show that oxidation of the polymer films occurs in the potential range of 180-190 mV and the reduction occurs in the range of 80-90 mV.

**Appendix:**

**I. Principle of cyclic voltammetry:**

We use a three electrode cell, containing a platinum plate (1cm×1cm) working electrode, a platinum wire auxiliary electrode and a calomel reference electrode.

The object of this set-up is to avoid the polarization of the SCE. In fact, the current passes from the working electrode to the auxiliary electrode. But it doesn't pass through the SCE (which is the case when we have a set-up with two electrodes) and it is recommended to use a millivoltmeter with a high impedance in order to avoid the current passage through the SCE. Moreover, this set-up allows to cancel the ohmic drop of the solution.

The electrodes are immersed in an electrolyte which ensures the conductivity by ionic migration.

The potentials are always taken in regard to the reference electrode.

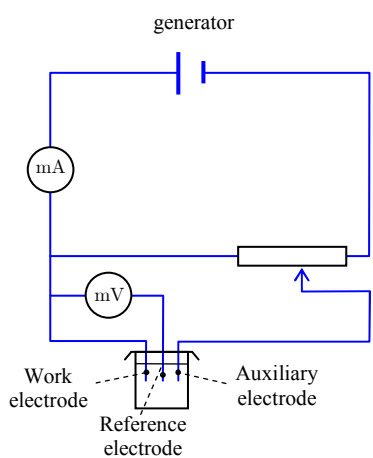

**II. How to get a voltamogram:**

a- <u>Potential variation</u>:

The potential of the working electrode varies linearly with the time from an initial value $E_i$ to a final value $E_f$. When the sweeping is finished, the direction of the potential variation is reversed.

The scheme below represents the variation of the potential versus the time.

The interval $[E_i, E_f]$ is chosen in a way that the solvent and the support electrolyte have no electroactivity.

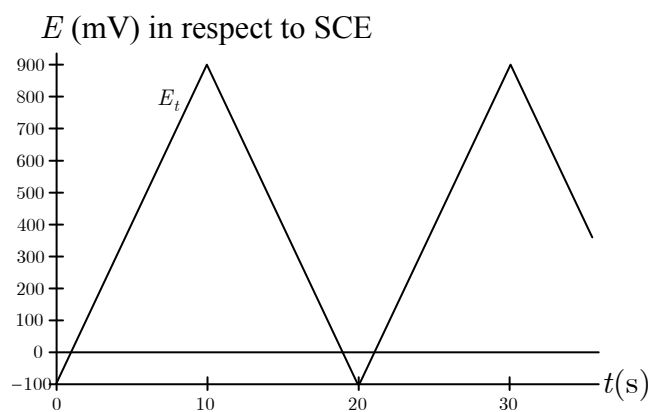

The slope of the signal is the scan rate. In this scheme it is $100\,\text{mV/s}$. We record the curve $i = f(E)$.

b- <u>Voltammogram interpretation</u>

An ideal voltammogram of a redox couple with a quick transfert of electrons is represented in the scheme below.

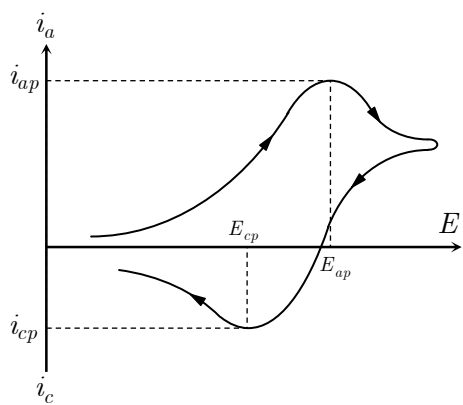

$i_{ap}$ : current of the anodic peak.

$i_{cp}$ : current of the cathodic peak.

$E_{ap}$ : potential of the anodic peak.

$E_{cp}$ : potential of the cathodic peak.

$$\Delta E_p = |E_{ap} - E_{cp}| = \frac{0.059}{n}$$

The potential $E_i$ is chosen in such a way that at the beginning of the sweeping no electrochemical reaction takes place. When this potential has a sufficient value, an oxidation reaction occurs and the intensity increases noticeably. The electroactive substance disappears quickly at the electrode surface.

This phenomenon generates a concentration gradient responsible of the diffusion. After a while, the arrival by diffusion of the electroactive substance can not compensate its disappearance due to the electrochemical reaction.

The current has a limited value. This is the diffusion plateau.

The same phenomenon happens in the reverse direction during the reverse sweeping. We notice the reduction of the oxidized specie.

If the couple is not reversible, but it represents however cathodic and anodic peaks, we say it's a quasi-reversible couple: $\Delta E_p \neq \dfrac{0.059}{n}$.

In some cases, we observe only one peak corresponding to a reduction or an oxidation. The system is considered irreversible.